\documentclass[11pt,twoside]{article}

\usepackage{asp2006}
\usepackage{epsf}
\usepackage{lscape}
\usepackage{graphicx}

\begin{document}
\title{Intermittent Activity of Jets in AGN}   

\author{Aneta Siemiginowska\altaffilmark{1}, Bo\.zena Czerny\altaffilmark{2}, 
Agnieszka Janiuk\altaffilmark{2}, {\L}ukasz Stawarz\altaffilmark{3}, 
Matteo Guainazzi\altaffilmark{4}, Annalisa Celotti\altaffilmark{5},
Giulia Migliori\altaffilmark{5} and  Olaf Tengstrand\altaffilmark{4}}

\altaffiltext{1}{Harvard-Smithsonian Center for Astrophysics, 60 Garden St, Cambridge, MA 02138, USA}
\altaffiltext{2}{Copernicus Astronomical Center, Bartycka 18, 00-716 Warsaw, Poland}
\altaffiltext{3}{KAVLI/Stanford University, Stanford, CA 94305, USA} 
\altaffiltext{4}{European Space Astronomy Center of ESA, P.O. Box 78, Vilanueva de la Canada, E-28691, Madrid, Spain}
\altaffiltext{5}{SISSA/ISAS, Via Beirut 2-4, I-34151, Trieste, Italy}

\begin{abstract} 

Large scale X-ray jets that extend to $>100$~kpc distances from the
host galaxy indicate the importance of jets interactions with the
environment on many different physical scales. Morphology of X-ray
clusters indicate that the radio-jet activity of a cD galaxy is
intermittent. This intermittency might be a result of a feedback
and/or interactions between galaxies within the cluster. Here we
consider the radiation pressure instability operating on short
timescales ($<10^5$ years) as the origin of the intermittent
behaviour. We test whether this instability can be responsible for
short ages ($ < 10^4$ years) of Compact Symmetric Objects measured by
hot spots propagation velocities in VLBI observations.  We model the
accretion disk evolution and constrain model parameters that may
explain the observed compact radio structures and over-abundance of
GPS sources. We also describe effects of consequent outbursts.

\end{abstract}


\section{Introduction}   

The idea of intermittency is not new. Some observational evidence was
given already in early 60-ties.
For example \cite{burbidge1965} suggested intermittent outbursts of
NGC1275 the cD galaxy in the center of Perseus A cluster.
\cite{kellerman1966} derived the intermittency timescales of
$\sim 10^4-10^6$~years required for the production of relativistic
particles responsible for the observed synchrotron spectra in radio
sources and quasars. Signatures of the past recurrent
activity in nuclei of normal galaxies were presented by
\cite{bailey1978}, but their paper was not really noticed and 
has only 24 citations to date.
\cite{shields1978} examined quasar models and suggested that their
accumulate the mass during quiescent periods and then through the
instability they transfer the mass onto a central black hole during
a short period of an outburst of the activity. These are just a few
examples of the AGN intermittency that has been considered since the
early days of studies of the nuclear emission in galaxies. It is now
that we are looking closely at this behaviour as it becomes evident
that it is an important component to our understanding of the
evolution of structures in the universe.  However, there are still
many open questions about the origin of the intermittent behaviour. Is
it related to the unsteady fuel supply, or accretion flow?  Are there
many quiescent and outburst phases?  What is the mechanism regulating
the intermittent behaviour?

Observations show a range of timescales for the AGN outbursts. Quasar
lifetimes estimated based on large samples of SDSS quasars are of
order of $10^7$~years \citep{martini2001}. Signatures of outbursts in
recent observations of X-ray clusters indicated similar timescales
\citep[see][for the review]{mcnamara2007}. However, episodes of activity on timescales
shorter than $<10^5$~years have been also observed for example in
compact radio sources
\citep{owsianik1998,reynolds1997} or as light echos in nearby galaxies
\citep{lintott2009}. 
In this review we will focus on the short timescales of the
intermittent jet activity. We discuss the radio source evolution,
observational evidence for the intermittent activity and present the
model for the origin and nature of the short term activity based on
the accretion disk physics.

\section{Jets and Compact Radio Sources}

Jets provide the evidence for energetic AGN outflows.  They highlight
the fact that the energy released in the nuclear region in the close
vicinity of a black hole can influence the environment at large
distances. They also trace the source age and activity timescales. In
many recently discovered X-ray jets associated with powerful quasars
the continuous X-ray jet emission extends to hundreds of kpc distances
from the core
\citep{siemi2002,sambruna2002,harris2006}. These jets are
straight, sometimes curved or bent, and have many knots. For example
in PKS~1127-145 jet associated with z=1.18 quasar the separation and
size of the knots may indicate separate outbursts of jet activity with
timescales $\sim10^5$~years \citep{siemi2007}.

Host galaxy scale jets, smaller than $<10$~kpc provide a direct
information about the jet interactions with the ISM and
feedback. Compact radio sources that are entirely contained within the
host galaxy represent the initial phase of radio source growth and
they are young.  The most compact Gigahertz Peaked Spectrum (GPS)
sources have linear radio sizes below $\sim 1$~kpc. Their age can be
probed directly by studying the expansion velocity of symmetric radio
structures and it is typically less than $10^3$~years
\citep{polatidis2003,gugliucci2005}. Compact Steep Spectrum (CSS) radio sources 
are slightly larger, but still contained within the host galaxy. Their
age is given by synchrotron ageing measurements and is smaller than
$10^5$~years \citep{murgia1999}. 

\section{Evolution of Radio Sources}


\begin{figure}[!ht]
\begin{center}
\includegraphics[scale=0.6]{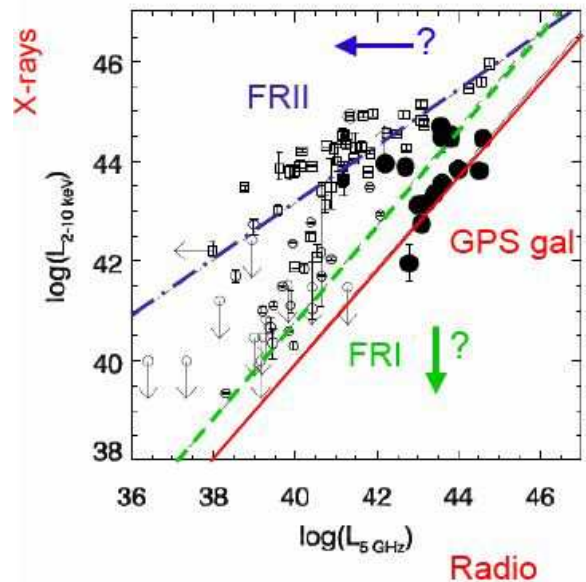}
\end{center}
\caption{Radio sources in the radio luminosity vs. X-ray luminosity
plane. FRI are marked with empty circles and FRII with squares.
Compact radio sources from \cite{tengstrand2009} are marked by large
black circles in the upper right-hand part of the diagram. The
regression lines for the FRI sources is marked with dashed line and
for FRII with dash-dot-dash lines. The solid line is the fit to the
compact sources with the lower X-ray luminosity. The evolutionary tracks
towards FRI and FRII sources are marked.}
\label{fig1}
\end{figure}


\cite{tengstrand2009} discuss the XMM-{\it Newton} and {\it Chandra} X-ray Observatory
observations of a complete sample of compact GPS galaxies.  All these
sources are unresolved in X-rays, but they are very powerful in both
the radio and X-rays bands. In Figure~\ref{fig1} we mark the location
of GPS galaxies in radio vs. X-ray luminosity plane. We also mark the
locations of the large scale FRI and FRII sources.  The GPS sources
cover an upper right corner of that diagram and the plotted regression
lines indicate possible evolutionary path for the GPS sources to grow
into large scale radio sources. However, it is unclear which way and
how the evolution of the GPS radio sources proceeds. Are they fade in
radio and X-rays simultaneously or maybe only the radio fades while
the X-ray emission remains unchanged indicating that the X-ray
emission process is independent of the radio one?  On the other hand
maybe the GPS sources remain in this part of the diagram due to
repetitive outbursts. \cite{odea1997} show that there is an
overabundance of compact radio sources, so they need to be short-lived
or intermittent rather then evolving in a self-similar manner to
explain their numbers.
\cite{reynolds1997} 
proposed a phenomenological model with the recurrent outbursts on
timescales between $\sim10^4-10^5$~years lasting for $\sim
3\times10^4$ years that fits the observed numbers of radio sources and their sizes.

There is additional growing evidence for the AGN intermittent
activity. Morphology of large radio galaxies with double-double or
triple-triple aligned structures show repetitions on scale of
$\sim10^5-10^6$ years \cite[see for example][]{brocksopp2007}).  Some
nearby radio galaxies show a younger radio structures embedded in a
relic radio halo \cite[e.g. 4C+29.30][]{jamrozy2007}.
Recent {\it Chandra} observations of X-ray clusters show signs of AGN
outbursts on timescales of $10^7$~year, e.g. Perseus A
\cite{fabian2003}, M87 \cite{forman2005}. However, the origin of intermittent activity
has not be determined so far. Mergers operate on long timescales and
can be important in X-ray clusters, especially at high
redshifts. Unstable fuel supply due to feedback may play a significant
role on the long timescales. Instabilities in the accretion flow
related to the accretion physics have been shown to operate on long
\citep[][]{shields1978,siemi1996,janiuk2004} and short timescales
\citep{janiuk2002,czerny2009} depending on the nature of the
instabilities.

\section{Outbursts of the Activity due to Accretion Disks Instabilities}

Accretion disk instabilities have been studied and shown to operate in
binary systems. The ionization instability that causes large amplitude
($\Delta L \sim 10^4$) outbursts in galactic binaries on timescales
between 1 and 1000 years may operate in accretion disks around
supermassive black holes. It can cause huge outbursts on timescale
$10^6-10^8$~years and influence the evolution and growth of a central
black hole \citep{siemi1996}. The radiation pressure instability
studied in microquasars causes moderate amplitude variability on short
timescales. The time-dependent disk models that include this
instability support the observed correlation between luminosity
variations and jet activity in microquasar \citep{fender2004}. In fact
the radiation pressure instability is the only quantitative mechanism
explaining the observed variability in GRS1915+105
\citep{janiuk2000,nayakshin2000,janiuk2002,merloni2006,janiuk2007}.
Scaling the observed outbursts with timescales of 100-2000~sec in this
$\sim 10 M_{\odot}$ system to 10$^9M_{\odot}$ galactic size black hole
gives the variability timescales between 300-6000 years.

\begin{figure}[!ht]
\begin{center}
\includegraphics[scale=0.48]{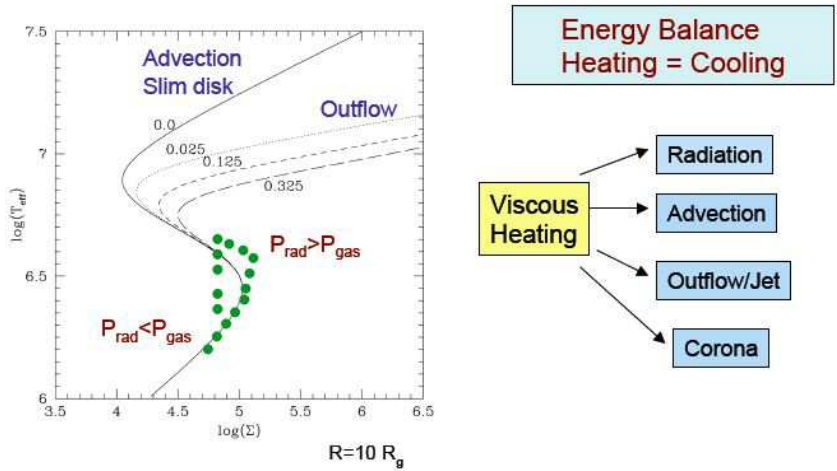}
\end{center}
\caption{ Surface density vs. effective temperature relation, e.g.
the local stability curve at 10$R_g$ for an accretion disk with the
viscosity scaled with gas and total pressure as
$\sqrt{P_{gas}P_{tot}}$. The viscous heating is balanced by cooling
through the processes listed on the right hand side.  The regions at
lower temperatures are dominated by the gas pressure, while the upper
branch becomes unstable when the radiation pressure dominates. Upper
curves shows different solutions that include parametrized outflow
stabilizing the disk. The top curve shows the solution for the
advective slim disk. Green solid qdots indicate solutions obtained during
time-dependent model calculation \citep[see][for
details]{janiuk2002}.}
\label{fig2}
\end{figure}


Figure~\ref{fig2} shows the standard stability curve in the surface
brightness vs. effective temperature plane with the stable solutions
for an $\alpha$ viscosity accretion disk.  In the region dominated by
radiation pressure the disk is unstable. The viscous heating cannot be
compensated by the radiative cooling and some additional cooling of
the disk in a form of an outflow, an advection towards the black hole
or the energy dissipation in the corona can stabilize the disk. The
solid dots show the evolutionary ``track'' of the local disk in the
unstable region. This model includes the outflow that transports away
the excess heating energy allowing the disk to transfer to a lower
temperature stable state
\citep{janiuk2002,czerny2009}.

\begin{figure}
\begin{center}
\includegraphics[scale=0.55]{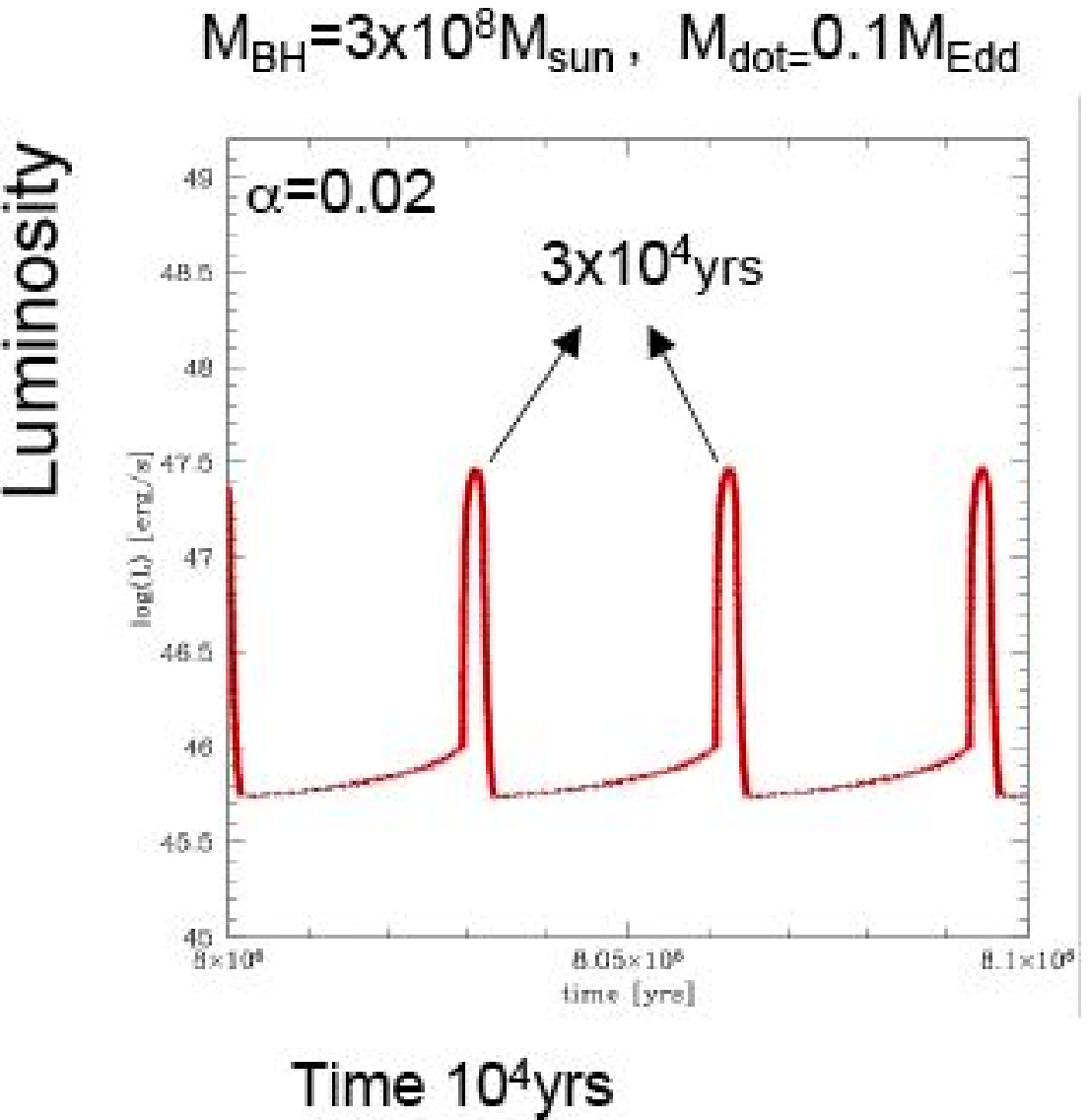}
\includegraphics[scale=0.48]{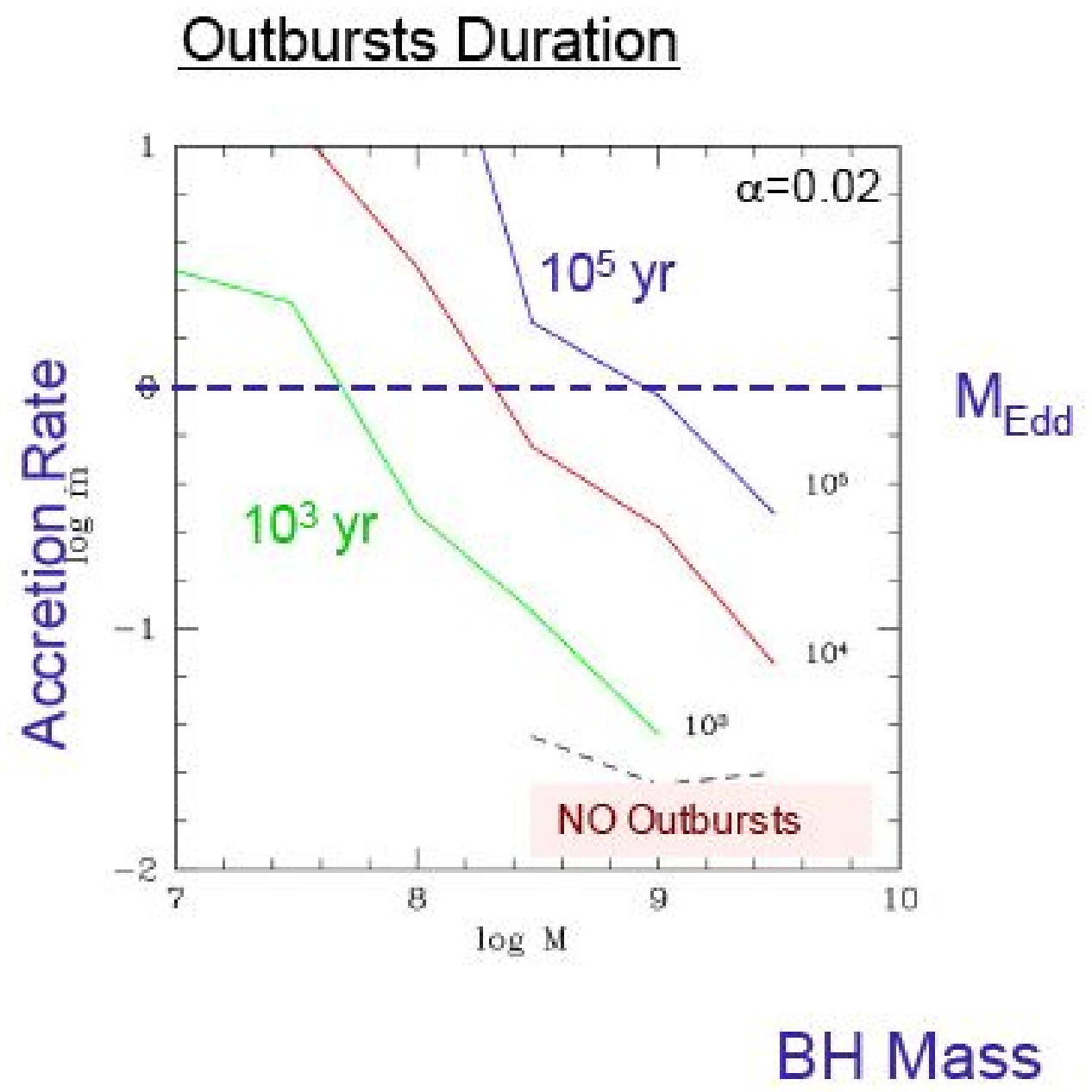}
\end{center}
\caption{{\bf Left:} Luminosity variations due to the radiation 
pressure instability in an accretion
disk around $M_{BH}=3\times10^8M_{\odot}$ and $\dot M =0.1 \dot
M_{Edd}$. {\bf Right:} Outburst durations represented as constant
lines in the black hole mass vs. accretion rate parameter
space. Outbursts lasting for $10^3$~years are too short for the
expansion of a radio source beyond its host galaxy.}
\label{fig3}
\end{figure}

Figure~\ref{fig3} shows the luminosity variations due to the radiation
pressure instability occurring in a disk around $M_{BH}=3\times10^8
M_{\odot}$ black hole with a steady flow of matter with the rate $\dot
M=0.1\dot M_{Edd}$, where $\dot M_{Edd}$ is the critical accretion
rate corresponding to the Eddington luminosity. The outbursts are
separated by 3$\times10^4$~years as required by
\citeauthor{reynolds1997} model. For a different black hole mass and
accretion rates the outbursts timescales and durations are
different. In general the effects of the instability depend on the
size of the unstable disk region which scales with accretion
rate. Also, there is a lower limit to the accretion rate for this
instability to operate because the radiation pressure instability
occurs when $P_{rad}>P_{gas}.$ In the systems with $M_{BH}>10^8
M_{\odot}$ and accretion rates below a few percent of the critical
Eddington accretion rates the instability does not occur and the disk
is stable.

We associate the outbursts caused by the radiation pressure
instability with the ejections of radio jets. The predicted timescales
for the outbursts durations and repetitions are in agreement with the
observations \citep{wu2009} of compact radio sources.

\section{Consequences}

There are several implications of the intermittent jet activity.

The repetitive outbursts have potentially a very strong impact on the
ISM. The repetitive shocks on timescales shorter than the relaxation
timescales may keep the ISM warm, and also more efficiently drive the
gas out of the host galaxy. 

There exists an intrinsic limit to the size of a radio source given by
the timescales.

Evolution of a radio source proceeds in a non self-similar way.  The
outbursts repeat regularly every $10^3-10^6$ years. The jet turns-off
between each outburst. Each radio structure may represent one
outbursts and a young source will indicate a new outburst. However,
the timescales for fading of a radio source are longer than the
separation timescales between the outbursts, and we should observe the
fossil radio structure in addition to the compact one.  10\% of GPS
sources have faint radio emission on large scales, that is typically
explained as a relic of the other active phase \citep{stanghellini2005}.

What is the radio source evolution between the outbursts?  The jet
turns-off between each outburst and then the pressure driven expansion
continues until the radio source energy is comparable to the thermal
energy of the heated medium. If the outburst lasts about $\sim
1000$~years then the radio source does not expand beyond the host
galaxy. The hot spots can only travel to a distance of $\sim 300$~pc
before the jet turns off and then the pressure driven expansion drives
the radio structure only to a distance of 3~kpc for the typical ISM
density and jet power. At this point the re-collapse of the radio
source starts and it continues for about a Myr if there is no another
outburst. In order for the source to escape its host galaxy the
outburst needs to last longer than $\sim 10^4$~years.

Re-collapse phase is typically longer than the repetition timescale and
a ``tunnel'' made by the jet will not close between outbursts.  It
means that the repetitive jets outbursts will propagate into a
rarified medium. As the timescales between outbursts are long we
expect to observe fading compact radio sources with no active nucleus.

For high accretion rates ($>0.02 \dot M_{Edd}$) the jet outbursts can
be governed by the radiation pressure instabilities. The observed
radio power suggests high accretion rates that are consistent with the
requirement for the instability to operate. Thus the number of compact
size radio sources with can be explained by the outbursts with
timescales consistent with the radiation pressure instability.

There are some open issues related to the theoretical understanding of
the disk instability.  The effects of the instability depend on the
disk viscosity. The standard \citeauthor*{shakura1973} disk is
unstable in the case of viscosity scaling with total pressure while it
is stable for the viscosity scaled with the gas pressure. In the case of
the MRI viscosity the local disk simulations support some scaling of the
effective torque with pressure.  However, the effect of the radiation
pressure instability is not clear in the global MHD simulations,
although very recent simulations by \cite{hirose2009} confirm the
existence of the instability.  Observationally the behaviour of
microquasars strongly supports the radiation pressure explanation for
the outbursts.

\section{Conclusions and Future Perspectives}

Observations indicate a complex behaviour of radio sources: continuous
jets, signatures of repetitive outbursts in separated radio
components, or the statistic of radio sources. The source complex
behaviour may reflect different regimes of the accretion flow that
depend on the black hole mass and accretion rate. We also note that
for some parameters the radio source may never leave the host galaxy.

Large samples of radio sources are needed for statistical studies. We
should be able to determine the number of sources, their lifetimes and
sizes. We also need more sources with measurements of their age as
well as indications for the intermittency in the sources radio
morphology. Such data should be available in the future with the new
radio surveys that probe fainter, low power compact sources that might
be in the fading phase.

\acknowledgements 

AS thanks the organizers for the invitation to the meeting.
This research is funded in part by NASA contract NAS8-39073. Partial
support for this work was provided by the NASA grants 
GO5-6113X, GO8-9125A and NNX07AQ55G.



\end{document}